\documentstyle[preprint,revtex]{aps}

\begin{document}
\thispagestyle{empty}
\font\fortssbx=cmssbx10 scaled \magstep2
\hbox{ 
\fortssbx University of Wisconsin - Madison} 
\vspace{.3in}
\hfill\vbox{\hbox{\bf MAD/PH/712}
	    \hbox{July 1992}}\par
\vspace{.2in}
\begin{title}
Test of an Equivalence Theorem at One-Loop
\end{title}
\author{M.~S.~Berger}
\begin{instit}
Physics Department, University of Wisconsin, Madison, WI 53706, USA\\
\end{instit}
\begin{abstract}
\nonum\section{abstract}
We show that the equivalence theorem approximating one-loop gauge sector
diagrams by including only Goldstone bosons in the loop
gives a remarkably poor approximation to
the amplitude for the decay $H\rightarrow \gamma \gamma $ and for the process
$\gamma \gamma \rightarrow HH$. At one loop, large logarithms can arise
that evade power counting arguments.
\end{abstract}

\newpage
The standard equivalence theorem\cite{ET} has become a popular method for
approximating difficult calculations. Amplitudes involving Goldstone bosons
substituted for external longitudinal gauge bosons are much easier
to calculate. It has been proven to all
orders in perturbation theory using power counting arguments at least for
external gauge bosons. Another possible application of this general concept
is to truncate one-loop calculations involving internal gauge bosons
to only those diagrams with just Goldstone bosons (no internal gauge bosons
or ghosts)\cite{BJ,JP}. This results in a separately finite and gauge
invariant sum and is clearly a simpler task than performing the full
calculation. In this short note we present examples where this
equivalence theorem (ET) performs
poorly. We find that the large logarithms that can appear at one-loop
destroy the approximation.
We choose processes that are absent at tree level and
first occur at one-loop. In this
way we are able to avoid any subtleties related to renormalization and
concentrate on the aspects that arise beyond tree level but are not
specifically related to any renormalization scheme. We do not argue that this
ET is invalid; rather the asymptotic approach to the limit can be gradual
and the predicted rates in physically interesting processes can be unreliable.

We discuss two processes in the Standard Model that are phenomenologically
interesting. One is well-known\cite{VVZS} and the
other has been considered relatively recently\cite{JP}.
First consider Higgs decay to two photons,
$H\rightarrow \gamma \gamma $. The full one-loop amplitude has been known for
some time\cite{VVZS}, and this process may serve as an interesting
theoretical laboratory for the ET. The ET can be employed to calculate the
Feynman diagrams involving the gauge boson sector in the loop.
Generic diagrams are shown in Figure 1. In the full calculation there are 26
diagrams, while only three diagrams contribute to the ET
approximation. We have calculated these diagrams in the Feynman gauge using
the symbolic manipulation programs FORM and MATHEMATICA.
We have used the algorithms for reducing one-loop integrals developed by
van Oldenborgh and Vermaseren\cite{loops}. This technique gives entirely
analytic expressions for the matrix elements. Gauge invariance is checked
analytically for the resulting expressions.

We find that the one-loop decay rate for the $W$ boson loops (not counting
fermion loops) determined by using the ET is
\begin{equation}
\Gamma ^{ET}_{}={{\alpha ^2G_FM_H^3}\over {16\sqrt{2}\pi ^3}}
\Big |1+2M_W^2C(p_1,p_2)\Big |^2 \;,
\end{equation}
The full calculation including all 26 diagrams yields\cite{VVZS}
\begin{equation}
\Gamma ^{FULL}_{}={{\alpha ^2G_FM_H^3}\over {16\sqrt{2}\pi ^3}}
\Big |\xi _1\left (1+2M_W^2C(p_1,p_2)\right )-8M_W^2C(p_1,p_2)\Big |^2 \;,
\end{equation}
where
\begin{equation}
\xi _1=\left [1+6{{M_W^2}\over {M_H^2}}\right ]\;,
\end{equation}
and $C(p_1,p_2)$ is the usual scalar three-point integral with two massless
external lines (see below) and can be expressed in logarithms alone,
\begin{equation}
C(p_1,p_2)={1\over {2M_H^2}}\ln ^2\left( {{-z}\over {1-z}}\right )\;,
\end{equation}
with
\begin{equation}
z={1\over 2}\left [ 1 + \sqrt {1-{{4M_W^2-i\epsilon }\over {M_H^2}}}\right ]\;.
\end{equation}
In the very small $M_W^2/M_H^2$ limit, $C(p_1,p_2)$ behaves as
$1/(2M_H^2)\ln ^2(M_W^2/M_H^2)$, and the subleading term in
$\Gamma ^{FULL}_{}$ cannot be neglected even for a heavy Higgs boson.
It is perfectly natural to expect logarithms and dilogarithms to arise in
one-loop graphs where integration over the loop momentum is performed.

In Figure 2 we plot the ratio $\Gamma ^{ET}_{}/\Gamma ^{FULL}_{}$ against the
Higgs mass $M_H$. Even for a
Higgs boson as heavy as 1 TeV, the decay rate has not begun to display the
asymptotic behavior prescibed by the ET. Eventually the ratio approaches one
but only for unrealistically large Higgs masses.

The argument presented so far might be considered only academic since the full
calculation for $H\rightarrow \gamma \gamma $ is well-known and easily
obtained, so we have also explored the effectiveness of the ET in the more
complicated process $\gamma \gamma \rightarrow HH$. This process has been
suggested as a possible method of measuring the triple-Higgs vertex in the
Standard Model. The ET calculation
of the $W$ boson loop contribution has been recently computed\cite{JP}. We have
performed the full one-loop calculation and find the ET calculation to
be inaccurate in the region of interest. There are 188 one-loop Feynman
diagrams in the full calculation compared to 22 in the ET approximation. See
Figure 3.

We have confidence in our result for the following reasons:
(1) We have checked analytically that $p_1^{\mu}T_{\mu \nu}=0$ and
$p_2^{\nu}T_{\mu \nu}=0$ where $T_{\mu \nu}$ is the polarization tensor for
$\gamma \gamma \rightarrow HH$. (2) We have used our program to reproduce
published helicity amplitudes for $gg \rightarrow HH$
(quark loop)\cite{gghh}, $gg \rightarrow ZZ$ (quark loop)\cite{ggzz}, and the
equivalence theorem part of $\gamma \gamma \rightarrow HH$\cite{JP}.
(3) The simple diagrams in Figure 3 (but not the boxes) were checked versus
the program FeynCalc/FeynArts\cite{FC}.

The helicity amplitudes can be expressed in a compact form using
the results obtained in the ET approximation in Ref.~\cite{JP}. We find
\begin{eqnarray}
{\cal M}_{++}^{FULL}=\xi _2{\cal M}_{++}^{ET,3a}&+&\xi _1{\cal M}_{++}^{ET,3b}
+2\Big (-YD(p_1,p_3,p_2)\nonumber \\
&&+2t_1C(p_1,p_3)+2u_1C(p_2,p_3)+6{{sM_H^2}
\over {s_1}}C(p_1,p_2)\Big ) \label{ppfull}\;, \\
{\cal M}_{+-}^{FULL}=\xi _3{\cal M}_{+-}^{ET,3a}
&+&2\Big (-YD(p_1,p_3,p_2)+t_1^2D(p_2,p_1,p_3)\nonumber \\
&&+u_1^2D(p_1,p_2,p_3)
+2t_1C(p_1,p_3)+2u_1C(p_2,p_3)+2sC(p_1,p_2)\Big )\;,\nonumber \\
\label{pmfull}
\end{eqnarray}
where
\begin{eqnarray}
\xi _2&=&\left [1-4{{M_W^2}\over {M_H^4}}\left (M_H^2-3M_W^2+s\right )
\right ]\;, \\
\xi _3&=&\left [1-4{{M_W^2}\over {M_H^4}}\left (M_H^2-3M_W^2-s\right )
\right ]\;,
\end{eqnarray}
and $s_1=s-M_H^2$, $t_1=t-M_H^2$, $u_1=u-M_H^2$, $Y=tu-M_H^4$.
The matrix elements ${\cal M}_{++}^{ET,3a}$, ${\cal M}_{++}^{ET,3b}$ and
${\cal M}_{+-}^{ET,3a}$ are those obtained in the ET approximation and given
in Ref.~\cite{JP} as ${\cal M}_0(box)$, ${\cal M}_0(triangle)$ and
$-{\cal M}_2(box)$ respectively (The minus sign is simply a matter of our
convention for the polarization vectors. The $M_H^2$ in the last line of
${\cal M}_2(box)$ should be $M_H^4$.).
The indices $3a$ and $3b$ refer to the diagrams in Figure 3.
The scalar triangle and box diagrams are defined as
\begin{equation}
\scriptstyle
C(p_1,p_2)={{1}\over {i\pi ^2}}\int d^4q{{1}\over
{(q^2-M_W^2)((q+p_1)^2-M_W^2)((q+p_1+p_2)^2-M_W^2)}}\;, \label{eq:tri}
\end{equation}
\begin{equation}
\scriptstyle
D(p_1,p_2,p_3)={{1}\over {i\pi ^2}}\int d^4q{{1}\over
{(q^2-M_W^2)((q+p_1)^2-M_W^2)((q+p_1+p_2)^2-M_W^2)((q+p_1+p_2+p_3)^2-M_W^2)}}
\;. \nonumber
\label{eq:box}
\end{equation}
The momenta of the incoming photons are $p_1$ and $p_2$ while the outgoing
Higgs bosons have momenta $p_3$ and $p_4$.
In the limit $M_W^2/M_H^2\rightarrow 0, M_W^2/s\rightarrow 0$,
the factors $\xi _1$, $\xi _2$ and $\xi _3$ approach one with only power law
corrections. On the other hand the additional terms do not become negligible
immediately.

We do not list separately  the contributions from the diagrams in Figure 3a
and Figure 3b because in the full calculation
they are no longer separately gauge invariant. The graphs
for $\gamma \gamma \rightarrow H$ shown in Figure 1 are certainly a gauge
invariant set, but once the Higgs is allowed off-shell as in Figure 3b,
a contribution from the graphs in Figure 3a must be included to obtain gauge
invariance. This is not the case for either the subset of diagrams in the ET or
for fermion loop contributions. Details of this calculation and issues of
phenomenological interest will be presented in a longer paper.

In Figure 4 we compare the cross sections given by the full
calculation and given by the ET. The two converge in the appropriate limit,
but this convergence is quite mild. The disagreement is most severe for
unequal photon helicities, $\lambda _1=- \lambda _2$.
As the center of mass energy increases the
approximation gets better, but even at $s=2$ TeV there is a substantial
discrepancy.

We have found that large logarithms that arise at one-loop limit the
effectiveness of this ET. We believe this behavior is a
general feature of such calculations, and one must be careful not to place
too much confidence in such ET calculations beyond the tree level. At tree
level with internal gauge boson lines, this type of logarithm is absent, and
the ET should converge quite
rapidly to the full result in the appropriate limit.
We have not specifically addressed the issue of one-loop diagrams with the
external longitudinal gauge bosons replaced with Goldstone bosons. We believe
large logarithms potentially plague these approximations as well.

\newpage
{\Large Figures}
\vskip 1in

{\bf Figure 1}: Generic diagrams in the $W$ boson loop contribution to
$H\rightarrow \gamma \gamma $. The loops consist of all possible combinations
of $W$ bosons, Goldstone bosons and ghosts. The number of nonzero diagrams is
shown, and the number of diagrams in the equivalence theorem calculation is
given in parentheses.

{\bf Figure 2}: Comparison of the full calculation to the equivalence theorem
approximation for $H\rightarrow \gamma \gamma $. The approximation only
becomes good for unrealistic Higgs masses.

{\bf Figure 3}: Generic diagrams in the $W$ boson loop contribution to
$\gamma \gamma \rightarrow HH$. The loops consist of all possible combinations
of $W$ bosons, Goldstone bosons and ghosts. The number of nonzero diagrams is
shown, and the number of diagrams in the equivalence theorem calculation is
given in parentheses.

{\bf Figure 4}: Comparison of the full calculation to the equivalence theorem
approximation for $\gamma \gamma \rightarrow HH$. The full calculation is
given by solid lines and the ET result by dashed lines.
(a) The cross sections for
constant center of mass energy $s=1$ TeV. (b) The cross sections for
constant center of mass energy $s=2$ TeV. (c) The cross section for $s=8M_H^2$
for the case where the two photons have unequal helicity.

\end{document}